\documentclass[twocolumn,showpacs,superscriptaddress,%
aps,prd,notitlepage,longbibliography,nofootinbib]{revtex4-1}

\usepackage{graphicx}
\usepackage{cancel}
\usepackage{amsmath}
\usepackage{amsfonts}
\usepackage{amssymb}
\usepackage{bbm}
\usepackage{verbatim}
\usepackage{maybemath}

\newcommand{\calF}{{\mathcal F}}
\newcommand{\calG}{{\mathcal G}}

\newcommand{\calO}{{\mathcal O}}

\def\calC{{\mathcal C}}
\def\calD{{\mathcal D}}

\def\calO{{\mathcal O}}

\def\calS{{\mathcal S}}

\def\bfM{{\bf M}}

\def\sfG{{\sf G}}

\newcommand{\cl}{{\mathrm{cl}}}
\newcommand{\LL}{{\mathrm{L}}}

\newcommand{\dd}{{\mathrm{d}}}

\def\tfrac12{{\textstyle\frac12}}

\def\dd{{\mathrm d}}
\def\ii{{\mathrm i}}
\def\ee{{\mathrm e}}
\def\cl{{\mathrm{cl}}}

\def\tfrac#1#2{ {\textstyle{\frac{#1}{#2}} } }

\begin{document}

\title{Instantons in $\maybebm{\phi^4}$ Theories: 
Transseries, Virial Theorems and Numerical Aspects}

\author{Ludovico T. Giorgini}

\affiliation{Nordita, Royal Institute of Technology and Stockholm University,
Stockholm 106 91, Sweden}

\author{Ulrich D. Jentschura}

\affiliation{Department of Physics, Missouri University of Science and
Technology, Rolla, Missouri 65409, USA}

\author{Enrico M. Malatesta}

\affiliation{Department of Computing Sciences, Bocconi University,
via Sarfatti 25, 20136 Milan, Italy}
\affiliation{Institute for Data Science and Analytics,
Bocconi University, 20136 Milano, Italy}

\author{Tommaso Rizzo}

\affiliation{Dipartimento di Fisica, Sapienza Universit\`a di Roma, P.le Aldo
Moro 5, 00185 Rome, Italy}

\affiliation{Institute of Complex Systems (ISC) - CNR, Rome unit, P.le A. Moro
5, 00185 Rome, Italy}

\author{Jean Zinn-Justin}

\affiliation{IRFU/CEA, Paris-Saclay, 91191 Gif-sur-Yvette Cedex, France}

\begin{abstract}
We discuss numerical aspects of instantons 
in two- and three-dimensional $\phi^4$ 
theories with an internal $O(N)$ symmetry group,
the so-called $N$-vector model.
Combining asymptotic transseries expansions for 
large argument with convergence acceleration 
techniques, we obtain high-precision values 
for certain integrals of the instanton that 
naturally occur in loop corrections around
instanton configurations. Knowledge of these
numerical properties are necessary in order to 
evaluate corrections to the large-order 
factorial growth of perturbation theory in $\phi^4$ theories.
The results contribute to the 
understanding of the mathematical structures underlying
the instanton configurations.
\end{abstract}

\maketitle

\tableofcontents

%
%
\section{Orientation}
\label{sec1}

The $N$-vector model (the self-interacting $\phi^4$ 
field theory in $D=2$ and $D=3$ dimensions) 
gives rise to instanton configurations,
whose structure is more complicated 
than the corresponding configurations in 
quantum mechanics (in one space dimension), which is equivalent
to a $D=1$ dimensional field theory
(see Figs.~2 and~3 of Ref.~\cite{JeSuZJ2009sigma}).
The instantons provide a nontrivial saddle point 
of the Euclidean action, about which we expand 
partition functions, and generating 
functions~\cite{BrPa1978,MaPaRi2017,GiEtAl2020,GiEtAl2022,GiEtAl2024ii}.
Instantons also constitute
fundamental objects in statistical 
and optimization problems possessing hard 
phases (see Refs.~\cite{Ma2023,MaKi2023,Ri2021}.
Here, we derive a semi-analytic representation which 
can be used to describe the instanton
uniformly over the radial variable, to 
a relative accuracy of $10^{-22}$ or better.

In one dimension (1D), one canonically identifies the argument of the
instanton as the Euclidean ``time'' $t$,
with the notion that $-\infty < t < \infty$
(see Ref.~\cite{JeSuZJ2009sigma}).
In 2D and 3D, this is not so easy,
because the angular symmetry dictates
that one should choose a radial variable.
The radial variable $r$, in turn, can
only take values in the range $0 < r < \infty$.
The connection to the 1D case~\cite{GiEtAl2020} is found if we consider
that in 1D, we can interpret the ``radial'' variable
with the $\mathbbm{Z}_2$ symmetry 
(positive and negative real numbers).
The surface area of the zero-dimensional
unit sphere embedded in one-dimensional space
is $2 \pi^{(D=1)/2}/\Gamma((D=1)/2) = 2$;
the result confirms the $\mathbbm{Z}_2$ symmetry
of the (analytically known) instantons in 
one-dimensional theories~\cite{JeSuZJ2009sigma,GiEtAl2020}.

In two-dimensional and three-dimensional $\phi^4$ theories,
the instanton is not known analytically.
Here, we aim to demonstrate that 
the analytic structure of the instanton is shown
to be linked to the concept of 
transseries and resurgent expansions
(see Refs.~\cite{Ph1989,CaNoPh1993,Bo1994,Ed2009,AnBaSc2019,Do2019,%
vSVo2022,DuHa2021,BoBr2022}). 
Specifically, we derive an asymptotic representation of the 
instanton, for large argument, in the form of 
a transseries (resurgent expansion)
in the variables $\chi = 1/r$ and $\exp(-1/\chi) = \exp(-r)$, 
where $r$ is the distance from the origin.
The transseries representation for large $r$
is complemented by a power-series representation 
for small $r$, which is augmented by Pad\'{e} approximants and
nonlinear sequence transformations to enhance its applicability 
for intermediate values of the radial variable.
The goal is to match the large-$r$ and small-$r$ 
representations at a suitable
intermediate transition value of the radial 
variable, to obtain a uniform, 
high-precision representation of the 
instanton in 2D and 3D. 

We organized the paper as follows.
Fundamentals of instantons in $\phi^4$ theories are
discussed in Sec.~\ref{sec2}.
The three-dimensional instanton in a 
three-dimensional $\phi^4$ theory is analyzed in 
Sec.~\ref{sec3}.
Our analysis of the instanton configuration 
in a two-dimensional field theory follows
in Sec.~\ref{sec4}.
Virial theorems and the asymptotic behavior of the 
instanton are discussed in Sec.~\ref{sec5}.
The high-precision evaluation of instanton integrals 
and of instanton actions is discussed in 
Sec.~\ref{sec6}. 
Conclusions are drawn in Sec.~\ref{sec7}.

%
%
\section{Fundamentals of Instantons in $\maybebm{\phi^4}$ Theories}
\label{sec2}

\subsection{Instanton Equations}

For the consideration of the instanton configuration,
it is sufficient to consider the $D$-dimensional scalar theory,
with the action
\begin{equation}
\label{ActionScalar}
S[\phi] = \int \dd^D x \, \left[
\frac12 \, \left( \vec\nabla \phi(\vec x) \right)^2 +
\frac12 \, \phi(\vec x)^2 +
\frac{g}{4} \, \phi(\vec x)^4 \right] \,,
\end{equation}
where $\vec x$ is a $D$-dimensional vector.
Consideration of the variation $\delta S[ \phi ] $
leads to the defining equation of the instanton,
\begin{equation}
\label{defeq_instanton}
- \vec\nabla^2 \phi_\cl(\vec x)  + 
\phi_\cl(\vec x) \, + g \, \phi_\cl(\vec x)^3  = 0  \,.
\end{equation}
Differentiation with respect to a coordinate leads to 
the equation of the zero mode $\partial_\mu \phi_\cl(\vec x)$,
\begin{equation}
\left( - \vec\nabla^2  + 
1 + 3 \, g \, \phi_\cl(\vec x)^2 \right)  \partial_\mu \phi_\cl(\vec x) = 0  \,.
\end{equation}
where $\mu = 1, \dots, D$.
In a quartic theory, the instanton solution
exists only for negative $g$, because the tunneling
can proceed only through a barrier.
Therefore, with the scaling
\begin{equation}
\label{defxi}
\phi_\cl(\vec x) = \sqrt{-\frac{1}{g}} \, \xi_\cl(\vec x) \,,
\end{equation}
the equations for the instanton and the zero mode are,
respectively,
\begin{align}
\label{eqinst}
\left( - \vec\nabla^2  + 
1 - \xi_\cl(\vec x)^2 \right) \xi_\cl(\vec x) =& \; 0  \,,
\\
\label{eqzero}
\left( - \vec\nabla^2  + 
1 - 3 \, \xi_\cl(\vec x)^2 \right)  \partial_\mu \xi_\cl(\vec x) =& \; 0  \,.
\end{align}
In a theory with an internal $O(N)$ 
symmetry group, one has the following instanton,
\begin{equation}
\underline{\phi}_\cl(\vec x) = 
\phi_\cl(\vec x) \, \underline{u} = 
\sqrt{-\frac{1}{g}} \, \xi_\cl(\vec x) \, \underline{u} \,,
\end{equation}
where vectors in the internal space are 
designated by underlining, and we can choose 
\begin{equation} 
\underline{u} = \{ 1, 0, \cdots, 0 \}^{\rm T} \,.
\end{equation}
As is evident from Figs.~2 and ~3 of Ref.~\cite{JeSuZJ2009sigma},
there is a sign ambiguity in the 
choice of the instanton, and the degeneracy 
under the operation $\xi_\cl \to -\xi_\cl$
needs to be taken into account when 
using dispersion relations.
Indeed, via dispersion relations, 
one can establish that the instanton action $A$,
defined via
\begin{equation}
\label{norm_action}
S[\phi_\cl] = -A/g \,,
\end{equation}
governs the large-order behavior of the perturbative 
coefficients $\sfG_K$ in the $K$th order 
of the expansion in $g$ of the $n$-point 
correlation functions in a $D$-dimensional 
$O(N)$ theory~\cite{GiEtAl2020}.
In the notation adopted in Eq.~(1.9) of Ref.~\cite{GiEtAl2020},
we have 
\begin{align}
\label{genexp}
\sfG_K = & \; 
\frac{c(N, D)}{\pi} \,
\left( \frac{1}{A} \right)^{(n+N+D-1)/2} \,
\left(-\frac{1}{A}\right)^K \, 
\\
& \; \times
\Gamma\left(K+  \frac{n + N + D - 1}{2} \right) \,
[ 1 + \calO(1/K)] \,.
\end{align}
Here, $c(N, D)$ is a constant coefficient
to be determined separately for each 
$N$ and $D$.

\begin{figure}[t!]
\begin{center}
\begin{minipage}{0.7\linewidth}
\begin{center}
\includegraphics[width=0.7\linewidth]{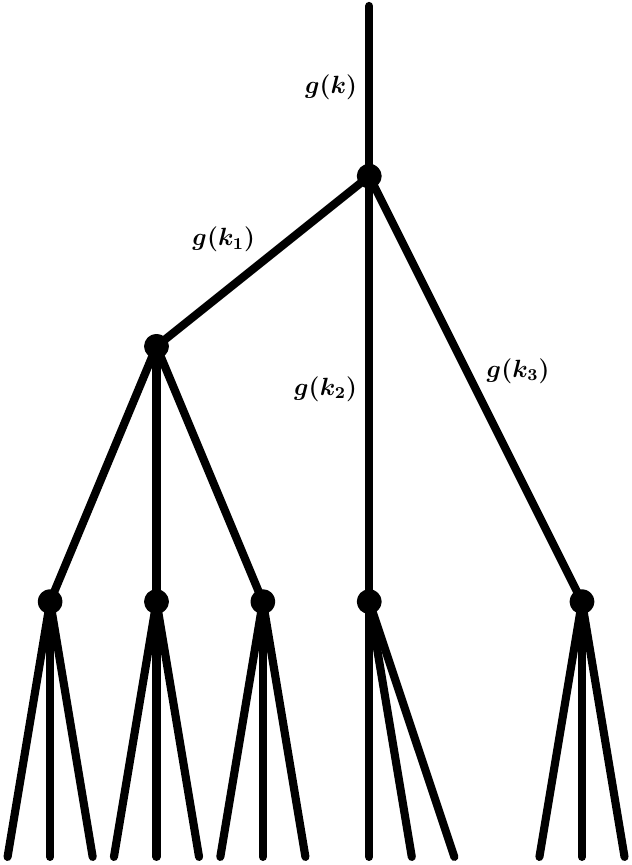}
\caption{\label{fig1} Tree diagram illustrating the
emergence of the instanton.}
\end{center}
\end{minipage}
\end{center}
\end{figure}

%
%
\subsection{Instantons and Large--Order Behavior}

The connection between instantons and large-order behavior is usually obtained
by saddle-point evaluations of contour integrals~\cite{ZJ2002}. In the
following, we will mention a less known derivation~\cite{Pa1977} that has the
advantage of being simpler and more intuitive. The basic idea is that
Feynman diagrams of the $\phi^4$ theory at large orders $K \gg 1$ are
essentially random regular graphs with connectivity~4 and size $K$. 
For a large number of vertices, it is known that random regular graphs have a locally
tree-like structure (with the size of the loops growing as $\log K$). This
allows us to write an iterative equation that turns out to be equivalent to the
instanton equation~\eqref{defeq_instanton}. The Feynman rules imply that there
is a factor $1/(\vec k_i^2 + m^2)$ for each line $i$ in the graph and a 
Dirac-$\delta$
function on each vertex ensuring the conservation of momentum. Invoking the
tree-like structure, one can then write the following equation (see also
Fig.~\ref{fig1})
\begin{multline}
g(\vec k) =\frac{1}{\vec k^2 + m^2}
\int \frac{\dd^D k_1}{(2 \pi)^D }
\int \frac{\dd^D k_2}{(2 \pi)^D }
\int \frac{\dd^D k_3}{(2 \pi)^D } \\ g(\vec k_1) g(\vec k_2)
g(\vec k_3)
  \delta^{(D)}(\vec k_1 + \vec k_2 + \vec k_3 - \vec k) \,.
\end{multline}
So, multiplying everything by $\vec k^2 + m^2$, the previous equation 
in real space is equivalent to 
\begin{equation}
(-\vec\nabla^2 + m^2) \, g(\vec x) = g^3(\vec x)\,,
\end{equation}
which is the instanton equation for the $\phi^4$ theory for $g(\vec x) =
\xi_\cl(\vec x)$.

Using a standard procedure (similar in spirit to the cavity method from 
spin-glass theory~\cite{MePaVi1986}), one can identify the action in
Eq.~\eqref{ActionScalar} as the Bethe free energy of the problem. This
allows us to derive in a simple way the instanton equation and the action; the
$1/K$ correction about the instanton~\cite{GiEtAl2020,GiEtAl2022}
corresponds to the $1/K$ finite size
correction to the Bethe lattice random graphs. 

%
%
\section{Three--Dimensional Instanton}
\label{sec3}

%
%
\subsection{Large Argument}
\label{3Dlarge}

We use the fact that
$\xi_\cl(\vec x) = \xi_\cl(|\vec x|) = \xi_\cl(r)$ is radially 
symmetric and this constitutes am ``$S$ state'' in the formalism 
adopted in atomic physics~\cite{ItZu1980,JeAd2022book}.
The equation fulfilled by the instanton 
$\xi_\cl(r) = \xi^{(3)}_\cl(r)$ (including the dimension
$D$ in the superscript) in three dimensions is
[see Eq.~\eqref{eqinst}], 
\begin{equation}
\label{trial}
- \frac{\partial^2}{\partial r^2} \xi^{(3)}_\cl(r) 
- \frac{2}{r} \frac{\partial}{\partial r} \xi^{(3)}_\cl(r) 
+ \xi^{(3)}_\cl(r) - [\xi^{(3)}_\cl(r)]^3 = 0 \,.
\end{equation}
We are attempting to find a systematic expansion 
of the solution of Eq.~\eqref{trial}, 
and do so for large argument $r \to \infty$ 
in the current section.
A remark might be in order.
Namely, linear second-order differential equations 
typically have solutions regular and irregular 
at infinite argument. The equations defining the 
instanton, by contrast, are highly nonlinear, and
hence this consideration does not apply. 
In fact, the asymptotics for large argument 
uniquely determine the instanton solution. 
In this context, it is instructive to 
recall~\cite{St2015} that 
the uniqueness of the solution for a 
{\em nonlinear} differential equation,
determined by a given initial condition or asymptotic behavior, 
constitutes a pivotal factor in the emergence 
of a range of complex phenomena, including
chaos. This uniqueness leads to behaviors which are sensitive to the small
variations in the initial
conditions~\cite{Lo1963Nonperiodic,Cl2021Nonlinear,Gl1994Stability}.
Furthermore, this intrinsic uniqueness in nonlinear differential equations is
analogous to the sensitive dependency on initial conditions observed in fluid
dynamics, particularly in the transition from laminar to turbulent flow, where
even minor perturbations can drastically alter the flow patterns, echoing the
underlying chaotic dynamics described in fluid mechanics
research~\cite{RuTa1971Turbulence}.

The instanton goes to zero as $r \to \infty$, and so 
one can neglect the term $[\xi^{(3)}_\cl(r)]^3 \ll \xi^{(3)}_\cl(r)$ 
in a first approximation.
Then, one obtains the relation
\begin{equation}
\label{firstorder}
- \frac{\partial^2}{\partial r^2} \xi^{(3)}_\cl(r)
- \frac{2}{r} \frac{\partial}{\partial r} \xi^{(3)}_\cl(r)
+ \xi^{(3)}_\cl(r) \approx 0 \,.
\end{equation}
Our {\em ansatz}
\begin{equation}
\label{ansatz}
\xi^{(3)}_\cl(r) = \frac{\exp(-r)}{r} \sum_{n=0}^\infty \frac{a_n}{r^n} \,,
\end{equation}
is a nonanalytic expansion in the variable $1/r$,
for large $r$.  In fact, in The variable $\chi = 1/r$, the 
expansion~\eqref{ansatz} constitutes a 
nonanalytic (resurgent, transseries) expansion in the variables
$\chi$ and $\exp(-1/\chi)$
(see Refs.~\cite{Ph1989,CaNoPh1993,Bo1994}),
and $\chi = 0$ becomes a singular point of the 
differential equation.  The importance of 
nonanalytic exponentials (resurgent expansions) in the solution of 
differential equations with singular points has been stressed in 
Ref.~\cite{Bo1994}.
The substitution $\xi^{(3)}_\cl(r) = g(r)/r$ takes 
Eq.~\eqref{firstorder} into the form
\begin{equation}
- \frac{\partial^2}{\partial r^2} g(r)
+ g(r) = 0 \,,
\qquad
g(r) = \calC \, \exp(-r) \,,
\end{equation}
for which the solution regular at infinity is just $\exp(-r)$.
Hence, the {\em ansatz}~\eqref{ansatz} collapses to a single term,
with $\calC$ being an overall constant, and reads
\begin{equation}
\xi^{(3)}_\cl(r) = \calC \, \frac{\exp(-r)}{r} \,,
\end{equation}
for which the approximate equality 
in Eq.~\eqref{firstorder} becomes an exact equality.
Here, $\calC$ is a coefficient which can be determined numerically.
A 60-figure result for $\calC$ is
\begin{align}
\label{numC3D}
\calC =& \;    2.712\,808\,360\,940\,844\,770\,465\,994\,573\,657 \nonumber\\
& \; \phantom{2.}808\,840\,265\,350\,950\,750\,281\,746\,458\,229(1) \,.
\end{align}
One can now approximate, in Eq.~\eqref{trial},
\begin{multline}
- \frac{\partial^2}{\partial r^2} \xi^{(3)}_\cl(r)
- \frac{2}{r} \frac{\partial}{\partial r} \xi^{(3)}_\cl(r)
+ \xi^{(3)}_\cl(r) - \xi^{(3)}_\cl(r)^3 
\\
\approx
- \frac{\partial^2}{\partial r^2} \xi^{(3)}_\cl(r)
- \frac{2}{r} \frac{\partial}{\partial r} \xi^{(3)}_\cl(r)
+ \xi^{(3)}_\cl(r) -  \frac{\exp(-3 r)}{r^3}  = 0 \,.
\end{multline}
The structure of this equation justifies the {\em ansatz}
\begin{equation}
\xi^{(3)}_\cl(r) = \calC \, \frac{\exp(-r)}{r} 
+ \frac{\exp(-3 r)}{r} \, \sum_{n=0}^\infty \frac{b_n}{r^n} + \dots \,.
\end{equation}
\begin{widetext}
Matching of the $b_n$ coefficients leads to the result,
\begin{equation}
\xi^{(3)}_\cl(r) = \calC \, \frac{\exp(-r)}{r}
- \calC^3 \, \frac{\exp(-3 r)}{8 r^3} \,
\left( 1 - \frac{3}{2 r} 
+ \frac{21}{8 \, r^2} - \frac{45}{8 r^3}
+ \frac{465}{32 \, r^4} - \frac{2835}{64 \, r^5}
+ \frac{40005}{256 \, r^6} + \calO(r^{-7}) \right) + \dots \,.
\end{equation}
This expression, cubed, generates terms proportional to 
$\left[ \frac{\exp(-r)}{r} + \frac{\exp(-3 r)}{r^3} \right]^3 
\to \left( \frac{\exp(-r)}{r} \right)^2 \times
\frac{\exp(-3r)}{r^3} = \frac{\exp(-5r)}{r^5}$.
Now, we enter with the {\em ansatz}
\begin{align}
\xi^{(3)}_\cl(r) =& \; \calC \, \frac{\exp(-r)}{r}
- \calC^3 \, \frac{\exp(-3 r)}{8 r^3} \,
\left( 1 - \frac{3}{2 r}
+ \frac{21}{8 \, r^2} - \frac{45}{8 r^3}
+ \frac{465}{32 \, r^4} - \frac{2835}{64 \, r^5}
+ \frac{40005}{256 \, r^6} + \calO(r^{-7}) \right) 
\nonumber\\
& \; + \frac{\exp(-5 r)}{r} \, \sum_{n=0}^\infty \frac{c_n}{r^n} + \dots \,,
\end{align}
again into Eq.~\eqref{trial}, match the coefficients $c_n$,  and find 
\begin{align}
\xi^{(3)}_\cl(r) =& \; \calC \, \frac{\exp(-r)}{r}
- \calC^3 \, \frac{\exp(-3 r)}{8 r^3} \,
\left( 1 - \frac{3}{2 r}
+ \frac{21}{8 \, r^2} - \frac{45}{8 r^3}
+ \frac{465}{32 \, r^4} - \frac{2835}{64 \, r^5}
+ \frac{40005}{256 \, r^6} + \calO(r^{-7}) \right)
\nonumber\\[0.1133ex]
& \; + \calC^5 \, \frac{\exp(-5 r)}{64 r^5} \, 
\left( 1 - \frac{19}{6 r}
+ \frac{151}{18 \, r^2} - \frac{815}{36 r^3}
+ \frac{56921}{864 \, r^4} - \frac{1094215}{5184 \, r^5}
+ \frac{2592553}{3456 \, r^6} + \calO(r^{-7}) \right) \,.
\end{align}
Finally, with the contribution 
of order $\exp(-7 r)$ included, we have
\begin{align}
\label{uptoseven_3D}
\xi^{(3)}_\cl(r) =& \; \calC \, \frac{\exp(-r)}{r}
- \calC^3 \, \frac{\exp(-3 r)}{8 r^3} \,
\left( 1 - \frac{3}{2 r}
+ \frac{21}{8 \, r^2} - \frac{45}{8 r^3}
+ \frac{465}{32 \, r^4} - \frac{2835}{64 \, r^5}
+ \frac{40005}{256 \, r^6} + \calO(r^{-7}) \right)
\nonumber\\[0.1133ex]
& \; + \calC^5 \, \frac{\exp(-5 r)}{64 r^5} \,
\left( 1 - \frac{19}{6 r}
+ \frac{151}{18 \, r^2} - \frac{815}{36 r^3}
+ \frac{56921}{864 \, r^4} - \frac{1094215}{5184 \, r^5}
+ \frac{2592553}{3456 \, r^6} + \calO(r^{-7}) \right)
\nonumber\\[0.1133ex]
& \; - \calC^7 \, \frac{\exp(-7 r)}{512 r^7} \,
\left( 1 - \frac{29}{6 r}
+ \frac{271}{16 \, r^2} - \frac{3943}{72 r^3}
+ \frac{614143}{3456 \, r^4} - \frac{8322275}{13824 \, r^5}
+ \frac{80215771}{36864 \, r^6} + \calO(r^{-7}) \right) \,.
\end{align}
For the term proportional to $\exp(-3 r)$, we find the 
compact formula,
\begin{equation}
\label{compact3D}
- \calC^3 \, \frac{\exp(-3 r)}{8 r^3} \,
\left( 1 - \frac{3}{2 r} + \frac{21}{8 \, r^2} - \frac{45}{8 r^3}
+ \frac{465}{32 \, r^4} + \calO(r^{-5}) \right) 
= - \calC^3 \, \frac{\exp(-r)}{r} \,
\left( \textrm{Ei}(-2r) - 2\exp(2r)\textrm{Ei}(-4r) \right) \,\,,
\end{equation}
where $\textrm{Ei}(r)$ the exponential integral function,
but we were unable to find general expressions for the 
terms in the series multiplying the exponential factors 
$\exp(-5r)$ and $\exp(-7 r)$. 
\end{widetext}

Transseries in the variable $\chi = 1/r$ have been encountered
in the study of anharmonic 
oscillators~\cite{ZJJe2004i,ZJJe2004ii,JeSuZJ2010,JeZJ2011}.
They have also been investigated 
mathematically~\cite{Ph1989,CaNoPh1993,Bo1994}.
We see that only ``odd-transseries''
orders of the form $\exp[-(2n+1)/\chi]$ contribute. The ``one-transseries'' contribution 
to the instanton wave function is found to read as
$\calC \, \chi \, \exp(-1/\chi)$, without correction terms.
The expansion~\eqref{uptoseven_3D} shows that the 
large-argument expansion of the instanton wave function $\xi^{(3)}_\cl(r)$
is determined by a single constant $\calC$, whose 
numerical value is given in Eq.~\eqref{numC3D}.

An inspection shows that 
the perturbative coefficients in the variable $\chi = 1/r$ grow
factorially. In order to match the resurgent expansion 
for large argument with the Taylor expansion for 
small argument $r$, we have calculated terms up to the 
13-instanton contribution and summed the series,
starting from large values of $r$, down to $r = 3$, 
\begin{equation}
\label{defximatch3D}
\xi^{(3)}_\cl(r = 3.0) = 0.045\,013\,219\,071\dots \,,
\end{equation}
where a more precise result for $\xi^{(3)}_\cl$ at the 
matching point is given in Eq.~\eqref{matchD30}.
In the summation process, we have used 
$[40/40]$--Pad\'{e} approximations~\cite{CaEtAl2007} in order 
to sum the divergent perturbative series decorating the 
instanton contributions of order $\exp(-n r)$, where
$n$ is an odd integer, and $77$-order Weniger--Levin
transformations~\cite{We1989}, in order to verify the 
accuracy of the result~\eqref{defximatch3D} in the intermediate region
near $r \approx 3.0$.

%
%
\subsection{Small Argument}
\label{3Dsmall}

We recall the equation fulfilled by the instanton 
[see Eq.~\eqref{trial}]
\begin{equation}
\label{trial2}
- \frac{\partial^2}{\partial r^2} \xi^{(3)}_\cl(r)
- \frac{2}{r} \frac{\partial}{\partial r} \xi^{(3)}_\cl(r)
+ \xi^{(3)}_\cl(r) - \xi^{(3)}_\cl(r)^3 = 0 \,.
\end{equation}
Plugging in a polynomial {\em ansatz} into Eq.~\eqref{trial2}, with 
\begin{equation}
\label{F_3D}
\xi^{(3)}_\cl(0) = \calF = 4.337\,387\,679\,976\dots 
\end{equation}
[see also Eq.~\eqref{xiD30}], one finds
\begin{align}
\label{FFser}
\xi^{(3)}_\cl(r) =& \; \calF + \frac16 \, (\calF - \calF^3) \, r^2
+ \frac{1}{120} (\calF - 4 \,\calF^3 + 3 \, \calF^5) \, r^4 
\nonumber\\[0.1133ex]
& \; + \frac{\calF - 17 \, \calF^3 + 35 \, \calF^5 - 19 \, \calF^7}{5040} \, r^4 
+ \calO(r^6) \,.
\end{align}
Only even powers of $r$ contribute.
Using computer algebra~\cite{Wo1999}, one can easily 
determine all coefficients up to order $r^{80}$,
and write
\begin{equation}
\xi^{(3)}_\cl(r) = \sum_{n=0}^\infty a_{2 n} \, r^{2 n} \,.
\end{equation}
A closer inspection reveals that 
the series of the $a_{2n}$ is factorially divergent and 
alternating.  Still, one can use summation techniques 
to confirm the result~\eqref{defximatch3D} 
at the matching point $\xi^{(3)}_\cl(r = 3.0)$
[see also Eq.~\eqref{matchD30}].
In the summation process, we have used
$[62/62]$--Pad\'{e} approximations~\cite{CaEtAl2007} in order 
to sum the divergent perturbative series at 
$r = 3$, or alternatively $117$-order Weniger--Levin
transformations~\cite{We1989}.
This leads to the desired accuracy in the intermediate region.
Improvements of the numerical accuracy 
are possible when one expands the instanton about 
additional reference points (e.g., where $r$ assumes
the value of a small integer) and concatenates the 
expansions in regions of overlap.

%
%
\section{Two--Dimensional Instanton}
\label{sec4}

%
%
\subsection{Large Argument}
\label{2Dlarge}

\begin{figure}[t!]
\begin{center}
\begin{minipage}{1.0\linewidth}
\begin{center}
\includegraphics[width=0.87\linewidth]{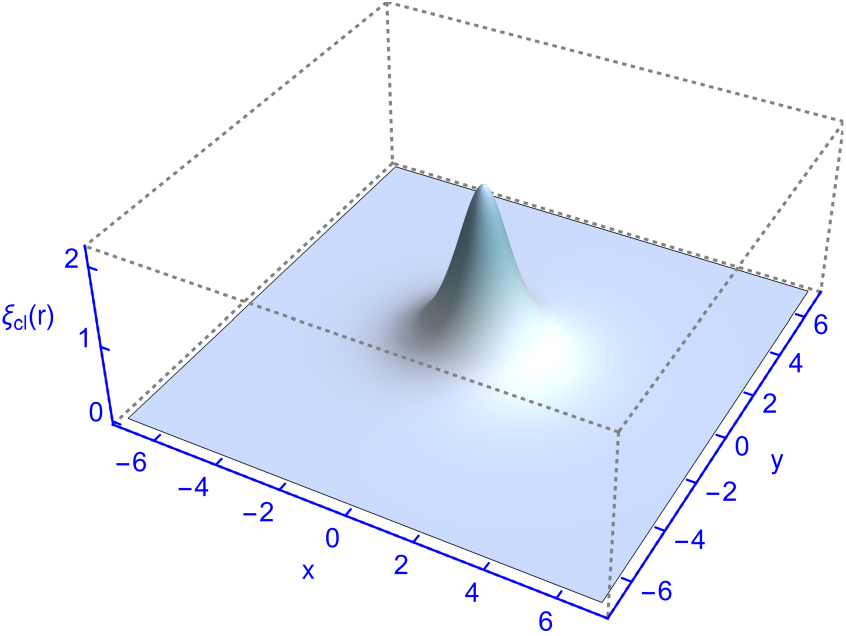}
\caption{\label{fig2}
The two-dimensional instanton $\xi_\cl(r) = \xi^{(2)}_\cl(r)$
is radially symmetric  Its value at the origin is
$\xi^{(2)}_\cl(0) = \calG = 2.206\,200\,864\,650\ldots$,
according to Eqs.~\eqref{F_2D} and~\eqref{xiD20}.}
\end{center}
\end{minipage}
\end{center}
\end{figure}

In two dimensions, the instanton is
equally radially symmetric (see Fig.~\ref{fig2}),
and we can write
$\xi^{(2)}_\cl(\vec x) = \xi^{(2)}_\cl(r)$.
The equation fulfilled by the instanton is
\begin{equation}
\label{trial3}
- \frac{\partial^2}{\partial r^2} \xi^{(2)}_\cl(r)
- \frac{1}{r} \frac{\partial}{\partial r} \xi^{(2)}_\cl(r)
+ \xi^{(2)}_\cl(r) - \xi^{(2)}_\cl(r)^3 = 0 \,.
\end{equation}
Just like in the three-dimensional case (see Sec.~\ref{3Dlarge}),
the instanton goes exponentially to zero as $r \to \infty$, and so
one can neglect the term $[\xi^{(2)}_\cl(r)]^3$ in a first approximation.
Then, one obtains the relation
[see also Eq.~\eqref{firstorder}]
\begin{equation}
- \frac{\partial^2}{\partial r^2} \xi^{(2)}_\cl(r)
- \frac{1}{r} \frac{\partial}{\partial r} \xi^{(2)}_\cl(r)
+ \xi^{(2)}_\cl(r) \approx 0 \,.
\end{equation}
By a similar analysis as described for the three-dimensional 
case, one obtains
\begin{widetext}
\begin{align}
\label{uptoseven_2D}
\xi^{(2)}_\cl(r) =& \; 
\calD \, \frac{\exp(-r)}{r^{1/2}} \,
\left( 1 - \frac{1}{8 r}
+ \frac{9}{128 \, r^2} - \frac{75}{1024 \, r^3}
+ \frac{3675}{32768 \, r^4} - \frac{59535}{262144 \, r^5}
+ \frac{2401245}{4194304 \, r^6} + \calO(r^{-7}) \right)
\nonumber\\[0.1133ex]
& \; - \calD^3 \, \frac{\exp(-3 r)}{8 r^{3/2}} \,
\left( 1 - \frac{9}{8 r}
+ \frac{213}{128 \, r^2} - \frac{3215}{1024 r^3}
+ \frac{238563}{32768 \, r^4} - \frac{5283711}{262144 \, r^5}
+ \frac{273186513}{4194304 \, r^6} + \calO(r^{-7}) \right)
\nonumber\\[0.1133ex]
& \; + \calD^5 \, \frac{\exp(-5 r)}{64 r^{5/2}} \,
\left( 1 - \frac{53}{24 r}
+ \frac{589}{128 \, r^2} - \frac{96193}{9216 r^3}
+ \frac{23551553}{884736\, r^4} - \frac{544320827}{7077888\, r^5}
+ \frac{85429251785}{339738624 \, r^6} + \calO(r^{-7}) \right) 
\nonumber\\[0.1133ex]
& \; - \calD^7 \, \frac{\exp(-7 r)}{512 r^{7/2}} \,
\left( 1 - \frac{79}{24 r}
+ \frac{10037}{1152 \, r^2} - \frac{629833}{27648 r^3}
+ \frac{55541473}{884736 \, r^4} - \frac{1326870785}{7077888 \, r^5}
+ \frac{23212812833}{37748736 \, r^6} + \calO(r^{-7}) \right) \,.
\end{align}
Just as in the three-dimensional case, we 
can find a compact expression for the 
leading term, which for $D=2$ is of order $\exp(-r)/\sqrt{r}$,
\begin{equation}
\label{compact2D}
\calD \, \frac{\exp(-r)}{r^{1/2}} \,
\left( 1 - \frac{1}{8 r}
+ \frac{9}{128 \, r^2} - \frac{75}{1024 \, r^3}
+ \frac{3675}{32768 \, r^4} + \calO(r^{-5}) \right)
= \ii \, \calD \, H^{(1)}_0(\ii r) \,.
\end{equation}
Here, $H^{(1)}_\alpha(r)$ is the Hankel function~\cite{AbSt1972} of the first kind 
of order $\alpha$.
However, we were unable to find general expressions for the
terms in the series multiplying the exponential factors
$\exp(-3r)$, $\exp(-5 r)$ and $\exp(-7 r)$.
\end{widetext}

For $D = 2$, one finds for the $\calD$ coefficient 
the following 60-figure result [cf.~Eq.~\eqref{numC3D}]
\begin{align}
\label{defC2D}
\calD =& \;    3.518\,062\,198\,025\,031\,180\,209\,129\,887\,741\nonumber\\
& \; \phantom{3.}356\,933\,215\,813\,390\,992\,384\,663\,366\,560(1)\,.
\end{align}
Furthermore, it is clear that
the perturbative coefficients in the variable $\chi = 1/r$ grow
very fast, and in fact, they grow factorially.
We have calculated terms up to the
contribution of order $\exp(-13 \,r)$ and summed the series,
starting from large values of $r$, down to
$r = 3$, with the result
\begin{equation}
\label{ref1_2D}
\xi^{(2)}_\cl(r = 3.0) = 0.097\,418\,218\,653\dots \,,
\end{equation}
where the most precise result for the
value of $\xi^{(2)}_\cl(r = 3.0)$ at the
matching point is given in Eq.~\eqref{matchD20}.
Just as in the three-dimensional case,
in the summation process, we have used
$[40/40]$--Pad\'{e} approximations in order
to sum the divergent perturbative series decorating the
instanton contributions of order $\exp(-n r)$, where
$n \leq 13$ is an odd integer, or $77$-order Weniger--Levin
transformations~\cite{We1989}, in order to achieve the
accuracy in the intermediate region.

%
%
\subsection{Small Argument}

We now need to repeat the analysis from Sec.~\ref{3Dsmall},
for the two-dimensional case.
Plugging in a polynomial {\em ansatz} into Eq.~\eqref{trial3}, 
\begin{equation}
\left( - \frac{\partial^2}{\partial r^2} -
\frac{1}{r} \frac{\partial}{\partial r}  + 1 
- \xi^{(2)}_\cl(r)^2 \right) \, \xi^{(2)}_\cl(r) = 0
\end{equation}
with
\begin{equation}
\label{F_2D}
\xi^{(2)}_\cl(0) = \calG = 2.206\,200\,864\,650\ldots \,,
\end{equation}
[see also Eq.~\eqref{xiD20}], one finds 
\begin{align}
\xi^{(2)}_\cl(r) =& \; \calG + \frac14 \, (\calG - \calG^3) \, r^2
+ \frac{1}{64} (\calG - 4 \,\calG^3 + 3 \, \calG^5) \, r^4
\nonumber\\[0.1133ex]
& \; + \frac{\calG - 19 \, \calG^3 + 39 \, \calG^5 - 21 \, \calG^7}{2304} \, r^4
+ \calO(r^6) \,.
\end{align}
Using computer algebra~\cite{Wo1999}, one can easily
determine all coefficients up to order $r^{80}$,
say, and write the divergent, asymptotic expansion
\begin{equation}
\xi^{(2)}_\cl(r) = \sum_{n=0}^\infty a_{2 n} \, r^{2 n} \,.
\end{equation}
Here, too, a closer inspection reveals that
the series is divergent, because of factorial
divergence of the magnitude of the
(alternating-in-sign) power series about $r = 0$.
One confirms the result given in Eq.~\eqref{ref1_2D}.
In the summation process, we have used
$[62/62]$--Pad\'{e} approximations in order
to sum the divergent perturbative series at
$r = 3$, or alternatively $117$-order Weniger--Levin
transformations~\cite{We1989}.
This yields the desired accuracy in the intermediate region.

%
%
\section{Instantons and Virial Theorems}
\label{sec5}

%
%
\subsection{Derivation of the Virial Theorems}
\label{sec5A}

For the following investigations, it is instrumental to derive several
relations fulfilled by the instanton; these constitute virial theorems.  We
consider the general action
\begin{equation}
\calS ( \phi ) = 
\int \dd^{D}x\left[
\frac12 \sum_\mu \left( \vec\nabla \phi(\vec x)\right)^2 + 
{\mathcal V}\bigl(\phi(\vec x)\bigr) \right]\,,
\end{equation}
where in the case of the action~\eqref{ActionScalar},
one has
\begin{equation}
{\mathcal V}\bigl(\phi(\vec x)\bigr) =
\frac12 \phi(\vec x) +
\frac14 g \, \phi^4(\vec x) \,.
\end{equation}
We assume that the field equation,
obtained by variational calculus, 
has a finite action solution $\phi_{\cl}(\vec x)$.
If the action $\calS ( \phi_{\cl} )$ is finite 
so is the action $\calS ( \phi_{\cl}, \lambda)$, 
obtained from $\calS ( \phi_{\cl} ) = 
\calS ( \phi_{\cl}, \lambda = 1)$ by the replacement 
$\phi(\vec x) \to \phi_{\cl}(\lambda \, \vec x)$.
If we change variables in the action,
setting $\lambda \vec x= \vec x'$, we find
\begin{multline}
\calS[ \phi_{\cl}, \lambda ] =
\lambda^{2-D} 
\int \dd^{D}x \, \left[ \frac12 \,
\sum_\mu \left(\vec \nabla \phi_{\cl}(\vec x) \right)^2 \right] 
\\
+ \lambda^{-D} \, 
\int \dd^{D}x \, {\mathcal V} \bigl(\phi_{\cl} (\vec x)\bigr) \,.
\end{multline}
Because $\phi_{\cl}(x)$ satisfies the field equation,
the variation of the action vanishes for $\lambda =1$,
i.e., we have the equation $\left. \frac{\dd}{\dd \lambda } 
\calS[ \phi_{\cl},\lambda ] \right|_{\lambda =1} = 0$.
\begin{multline}
\int \dd^{D}x' \,
\left[ \frac{D-2}{2} \left(\vec\nabla \phi_{\cl}(\vec x') \right)^2 
+ D \, {\mathcal V} \bigl(\phi_{\cl}(\vec x')\bigr) 
\right] =0\,.
\end{multline}

This relation allows us to express
the kinetic term [integral of $(\vec\nabla \phi_{\cl} )^2$]
in terms of the potential term
(integral of ${\mathcal V}(\phi_{\cl})$ and vice versa.
The classical action $\calS[\phi_{\cl}]$ 
can thus be expressed in terms of the kinetic term only:
\begin{equation}
\label{rel1}
\calS[ \phi_{\cl} ] = 
\frac{1}{D} \; \int \dd^{D}x \,
\left(\vec\nabla \phi_{\cl}(\vec x) \right)^2 \,,
\end{equation}
a form that shows that $ \calS ( \phi_{\cl} )$ is always positive. 
The second derivative of $ \calS ( \phi_{\cl}  ,\lambda ) $
reads as 
\begin{equation}
\label{eDerickb}
\left. 
\frac{\dd^2}{\left( \dd  \lambda \right)^2}
\calS[ \phi_{c}, \lambda ] \right|_{\lambda =1}
= \left(2-D \right) \,
\int \left[ \partial_{\mu}\phi_{c} (x) \right]^2 \dd^{D}x\,. 
\end{equation}
For $D\ge 2$, this result shows that  the solution is not a local minimum of
the action and, thus, the so-called longitudinal 
fluctuation operator ${\bfM}_\LL$, defined as 
\begin{equation}
\label{defM}
{\bfM}_\LL(\vec x,\vec x')= \left. 
\frac{ \delta^2\calS }{ 
\delta \phi(\vec x) \, \delta \phi(\vec x' )}
\right|_{\phi = \phi_\cl} \,,
\end{equation}
has at least one negative eigenvalue~\cite{JeZJ2011,GiEtAl2020}.

In the example of potentials of special form 
\begin{equation}
{\mathcal V}(\phi) = \frac12 \, \phi^2 + \frac14 \, g \, \phi^M \,, 
\end{equation}
one can derive an additional relation. 
If the action $\calS[\phi_{\cl}]$ is
finite, so is the following action 
obtained by the replacement
$\phi_\cl \to \lambda \, \phi_\cl$,
\begin{multline}
\calS[\lambda \phi_{\cl}] =
\lambda^2 \, \int \dd^{D}x \,
\frac12 \left[ \left(\vec\nabla \phi_{\cl}(\vec x) \right)^2 +
\phi_{\cl}^2(\vec x) \right] 
\\
+ \frac14 \, \lambda^M \, g\int \dd^{D}x \, \phi_{\cl}^M(\vec x) \,.
\end{multline}
Again,  if $\phi_{\cl}$ is the 
instanton solution, then the derivative with 
respect to $\lambda$ must vanish for $\lambda = 1$.
One obtains further relations, in addition to~\eqref{rel1},
\begin{multline}
\label{rel2a}
\calS ( \phi_{\cl} ) =
-\frac{g}{8} (M - 2) \, \int\dd^D x\,
\phi^M_{\cl}(\vec x) 
\\
\mathop{=}^{M=4} -\frac{g}{4}  \int\dd^D x\,
\phi^4_{\cl}(\vec x) 
\qquad (M=4) \,.
\end{multline}
This relation is consistent with the 
fact that the instanton exists only for negative $g$.
Thus, one can express the instanton action as follows,
\begin{multline}
\label{rel2b}
\calS ( \phi_{\cl} ) =
\frac{M-2}{ 2D-M(D-2)} \, 
\int\dd^D x\,\phi_{\cl}^2(x) 
\\
\mathop{=}^{M=4} \frac{1}{ 4 - D } \,
\int\dd^D x\,\phi_{\cl}^2(x) \,.
\end{multline}
We have used Eqs.~\eqref{rel1} and~\eqref{rel2a}.
In particular, Eqs.~\eqref{rel2a} and~\eqref{rel2b} 
are consistent only if the 
denominator in the expression on the right-hand side 
of Eq.~\eqref{rel2b} is positive, which implies 
\begin{equation}
M \le \frac{2D}{D-2} \,, 
\end{equation}
and, therefore,
the field theory must be super-renormalizable or at least renormalizable. At
the special dimension $D=2M/(M-2)$, where the theory is renormalizable, 
one finds the paradoxical result
$\int\dd^D x\,\phi_{\cl}^2(x)=0$. This implies that 
only the massless equation (where the coefficient of 
$\phi^2$ in the original action vanishes),
has instanton solutions (see also Appendix~\ref{appa}).  Finally, one
verifies that the second derivative at $\lambda=1$ is negative, confirming the
existence of a negative eigenvalue of $\bfM_\LL$ for all dimensions.  

%
%
\subsection{Summary of the Virial Theorems}

We summarize. From Eqs.~\eqref{rel1},~\eqref{rel2a}
and~\eqref{rel2b}, we have for the 
$\phi^4$ theory with the action~\eqref{ActionScalar},
\begin{multline}
\label{breakdown}
\calS ( \phi_{\cl} ) 
= \frac{1}{D} \; \int \dd^{D}x \,
\left(\vec\nabla \phi_{\cl}(\vec x) \right)^2 
= -\frac{g}{4}  \int\dd^D x\, \phi^4_{\cl}(\vec x)
\\
= \frac{1}{ 4 - D } \, \int\dd^D x\,\phi_{\cl}^2(\vec x) \,.
\end{multline}
With the scaling given by Eqs.~\eqref{defxi},
the action of the instanton becomes ($g < 0$),
\begin{subequations}
\label{eDericfiv}
\begin{align}
\label{derrick_a}
\phi_{\cl}(\vec x) =& \; \sqrt{ - \frac{1}{g} } \, \xi_{\cl}(r) \,,
\quad
r = | \vec x | \,,
\quad
\calS (\phi_{\cl}) = -\frac{A}{g} > 0 \,.
\end{align}
We have three equivalent representations
of the action $A$,
\begin{multline}
\label{derrick_b}
A = \frac{1}{D} \,
\int \dd^D x \left( \vec\nabla \xi_{\cl}(r) \right)^2 =
\frac14 \, \int \dd^D x \, \xi^4_{\cl}(r) \\
= \frac{1}{4-D} \, \int \dd^D x\, \xi^2_{\cl}(r) \,.
\end{multline}
\end{subequations}
Using the radial symmetry of the solution, we can establish that
\begin{equation}
\label{eAclassic} 
A = \frac{\Omega_D}{D} \; 
\int_0^\infty \dd r \, r^{D-1} \, [ \xi'_{\cl}(r) ]^2  \,,
\end{equation}
where $\Omega_D = 2\pi^{D/2} / \Gamma(\tfrac12 D)$
is the generalized surface of the $(D-1)$-dimensional 
unit sphere embedded in $D$-dimensional space.

%
%
\subsection{Asymptotic Behavior}

The asymptotic behavior of the radial instanton 
equation
\begin{equation}
\left[ - \left( \frac{\dd}{\dd r} \right)^2 -
\frac{D-1}{r} \frac{\dd}{\dd r} 
+ 1 \right] \xi_{\cl}(r) - \xi_{\cl}^3(r) = 0
\end{equation}
is of interest for large $r$.
Asymptotically, one can show that, for $r\to\infty$,
\begin{equation}
\xi_\cl(r)= {C\sqrt{\frac{2}{\pi}}} \; r^{1-D/2}K_{D/2-1}(r) + \calO(\ee^{-3r})
\end{equation}
where $K_\nu$ is a modified Bessel function of the second kind normalized such
that $K_\nu(r) \sim \sqrt{\pi/(2r)} \; \ee^{-r}$ for large $r$.
This would mean that 
\begin{equation}
\xi^{(D)}_\cl(r) \propto
{\sqrt{\frac{2}{\pi}}} \; r^{1-D/2} \sqrt{\frac{\pi}{2r}} \; \ee^{-r} 
= \frac{\ee^{-r}}{r^{D/2-1/2}} \,.
\end{equation}
Our formulas~\eqref{uptoseven_3D} and~\eqref{uptoseven_2D}
confirm this asymptotic behavior.

%
%
\section{Instanton Integrals}
\label{sec6}

We give a collection of numerical results for 
integrals of the instanton in quartic theories,
with enhanced accuracy.
Our aim is to give, for 2D and 3D, 
results approaching the
realm of applicability of the PSLQ
algorithm~\cite{FeBa1992,BaPl1997,FeBaAr1999,BaBr2001}
which is designed to search for analytic expressions
of integrals in terms of known constants.
First, for completeness,
in one dimension, we recall that the instanton solution 
is~\cite{JeSuZJ2009sigma}
\begin{equation}
\xi_\cl(r) = \frac{2}{\sqrt{ \cosh(r) + 1}} \,,
\qquad
\xi_\cl^{(D=1)}(r=0) = \sqrt{2} \,,
\end{equation}
where the one-dimensional action is
\begin{equation}
S[\phi] = 2 \int_0^\infty \dd r \, \left[
\frac12 \, \left( \partial_r \phi(r) \right)^2 +
\frac12 \, \phi(r)^2 +
\frac{g}{4} \, \phi(r)^4 \right] \,.
\end{equation}
The prefactor $2$ reflects on the angular factor 
$\Omega_D = 2\pi^{D/2} / \Gamma(\tfrac12 D)$ which 
evaluates to $2$ for $D=1$.
The prefactor matters because the instanton action 
is normalized to $S[\phi_\cl(r)] = -A/g$,
according to Eq.~\eqref{norm_action}.
Analytically known instantons 
in a four-dimensional $\phi^4$ theory 
and in a six-dimensional $\phi^3$ theory are
given in Appendixes~\ref{appa} and~\ref{appb}, respectively.

\begin{widetext}

We have noticed that instanton solutions
are determined by the value at the origin,
given in Eqs.~\eqref{F_3D} and~\eqref{F_2D}.
Numerically more accurate results can be obtained
via convergence acceleration algorithms, 
starting from a linear lattice of radial 
coordinates~\cite{JeGi2024}.
Explicit results, to 78~decimal figures, are given as follows,
\begin{subequations}
\begin{align}
\label{xiD10}
\xi_\cl^{(D=1)}(r=0) =& \; \sqrt{2} \,,
\\[0.1133ex]
\label{xiD20}
\xi^{(D=2)}_\cl(r=0) = \calG =& \;
     2.206\,200\,864\,650\,746\,074\,783\,634\,064\,578\,940\,196\,610\nonumber\\
& \; \phantom{2.}274\,520\,602\,192\,125\,757\,262\,456\,450\,184\,032\,518\,642(1)\,,
\\[0.1133ex]
\label{xiD30}
\xi_\cl^{(D=3)}(r=0) = \calF = & \;
     4.337\,387\,679\,976\,994\,356\,522\,109\,173\,841\,761\,465\,745\nonumber\\
& \; \phantom{4.}284\,082\,970\,785\,762\,761\,882\,558\,415\,947\,364\,399\,341(1)\,.
\end{align}
\end{subequations}
The values at the matching point $r = 3$ are interesting
for $D=2$ and $D = 3$,
\begin{subequations}
\begin{align}
\label{matchD20}
\xi^{(D=2)}_\cl(r=3.0) =& \; 
   0.097\,418\,218\,653\,642\,217\,741\,513\,024\,960\,584\,546\,095\nonumber\\
& \;  \phantom{0.}157\,618\,276\,680\,772\,556\,932\,915\,093\,354\,850\,219\,044(1)\,.
\\[0.1133ex]
\label{matchD30}
\xi_\cl^{(D=3)}(r=3.0) =& \; 
   0.045\,013\,219\,071\,010\,523\,997\,989\,047\,723\,112\,322\,109\nonumber\\
& \; \phantom{0.}014\,075\,244\,317\,789\,103\,014\,970\,206\,885\,072\,459\,490(1)\,.
\end{align}
\end{subequations}
Numerical results for the instanton action $A$ are
\begin{subequations}
\begin{align}
\label{AD1}
A(D=1) =& \; 4/3 \,,
\\[0.1133ex]
\label{AD2}
A(D=2) =& \; 
  \phantom{1}5.850\,448\,262\,279\,826\,939\,326\,986\,338\,934\,453\,868\,499\nonumber\\
& \;  \phantom{18.}064\,115\,959\,470\,267\,545\,644\,043\,014\,800\,957\,116\,007(1)\,.
\\[0.1133ex]
\label{AD3}
A(D=3) =& \; 
  18.897\,251\,302\,546\,190\,505\,297\,247\,993\,763\,227\,763\,807\nonumber\\
& \; \phantom{18.}178\,891\,316\,289\,857\,028\,151\,589\,245\,449\,182\,127\,167(1)\,.
\end{align}
\end{subequations}
These results are essential for large-order perturbation
theory [see Eq.~\eqref{genexp}].
For what follows, it is convenient to introduce the notation
\begin{equation}
\label{einstfixIn}
I_n=\int \dd^D x \, [\xi_\cl(r)]^n \,,
\qquad
I_2 = \frac{4-D}{4} \, I_4 \,, 
\qquad
I_4 = \frac{4}{D} \, 
\int \dd^D x \, [ \vec\nabla \xi_\cl(r) ]^2 \,,
\qquad
A = \frac14 I_4 \,,
\end{equation}
where we recall that the (generalized) surface area of the 
$(D-1)$-dimensional unit sphere, 
embedded in $D$ dimensions, is 
$\Omega_D = 2 \pi^{D/2} / \Gamma(D/2)$.
Results for $I_2$ and $I_4$ follow from the above results
for the instanton action $A$.
Results for $I_3$ and $I_6$ are given as follows,
\begin{subequations}
\label{integralsI}
\begin{align}
I_3(D=1)   =& \; \sqrt{2} \, \pi \,,
\qquad
I_6(D=1)   = 128/15 \,,
\\[2ex]
I_3(D=2)   =& \; \phantom{6}
 15.109\,669\,726\,889\,195\,199\,613\,754\,001\,702\,125\,888\,865 \nonumber\\
& \;    \phantom{615.}874\,563\,104\,430\,202\,476\,703\,241\,753\,965\,063\,516\,331(1) \,.
\\[2ex]
I_3(D=3)   =& \; \phantom{6}
   31.691\,521\,838\,323\,486\,451\,591\,907\,257\,120\,270\,170\,457 \nonumber\\
& \;    \phantom{631.}351\,985\,790\,745\,758\,122\,769\,708\,466\,866\,938\,412\,287(1) \,.
\\[2ex]
I_6(D=2)   =& \; \phantom{6}
  71.080\,171\,542\,041\,917\,440\,792\,898\,285\,323\,353\,751\,992 \nonumber\\
& \;    \phantom{615.}546\,589\,483\,197\,579\,092\,397\,461\,668\,330\,495\,246\,091(1) \,.
\\[2ex]
I_6(D=3)   =& \; 
  659.868\,351\,544\,567\,238\,188\,639\,540\,582\,544\,719\,267\,515 \nonumber\\
& \; \phantom{615.}748\,898\,263\,457\,303\,680\,826\,218\,470\,215\,371\,739\,493(1) \,.
\end{align}
\end{subequations}

\end{widetext}

%
%
\section{Conclusions}
\label{sec7}

We have analyzed the properties of 
instanton solutions in $O(N)$-symmetric
quartic field theories in $D=2$ and
$D=3$ dimensions.
The basic formulation for the quartic
instanton has been given in in Sec.~\ref{sec2}.
We concentrate on the three-dimensional
instanton ($D=3$) in Sec.~\ref{sec3},
which is phenomenologically the most
interesting case.
We derive asymptotic expansions for 
large argument in the form of a transseries 
[Eq.~\eqref{uptoseven_3D}] and in the form of an
asymptotic power series [Eq.~\eqref{FFser}]
for small argument.
The quartic instanton in $D=2$ is discussed in Sec.~\ref{sec4}.
Virial theorems are derived in Sec.~\ref{sec5}.
Instanton integrals are given in Sec.~\ref{sec6},
with a precision approaching the 
realm of applicability of the PSLQ 
algorithm~\cite{FeBa1992,BaPl1997,FeBaAr1999,BaBr2001}
which is designed to search for analytic expressions
of integrals in terms of known constants 
such as the Euler constant $\gamma_E = 0.57721\dots$,
various Riemann zeta functions, powers of $\pi$,
and multiplicative combinations of these constants.
We can report that we have carried out a limited set of searches 
with the same constants that were used in 
Eq.~(A11) of Ref.~\cite{GiEtAl2020} without success.
A more detailed search might constituent a possible
direction for the future.

In a quartic theory, the instanton solution
exists only for negative $g$, because the tunneling 
can proceed only through a barrier, and the latter exists 
only negative coupling $g < 0$. The imaginary 
part of the partition function,
and of correlation functions, obtained by expanding 
about the instanton solution, is 
proportional (see Ref.~\cite{GiEtAl2020})
to $\exp( - (-A/g) ) = \exp(A/g)$,
where $\calS[\phi_\cl] = -A/g$
and $A$ is given for $D=2$ in Eq.~\eqref{AD2}
for $D=3$ in Eq.~\eqref{AD3}.
The instanton action $A$ universally enters
large-order formulas for the perturbative
coefficients of Green functions [see Eq.~\eqref{genexp}].

Our calculations suggest that instanton solutions
in quartic theories cannot be expressed 
in closed analytic form, except for the 
case $D=1$. We also note that the general
properties of the instanton
in massive theories are valid 
only for dimensions $D < 4$.
The dimension 4 is singular, 
as is evident, e.g., from Eq.~\eqref{breakdown}.
The cases of a massless quartic theory 
in four dimensions, and of a cubic theory 
in six dimensions, are treated in Appendixes~\ref{appa} 
and~\ref{appb}.

Our investigations indicate that,
with the exception of known singular cases
(see Appendixes~\ref{appa} and~\ref{appb}),
instanton configurations cannot be 
calculated analytically for general
field theories, notably, for the 
two- and three-dimensional $\phi^4$ theories. 
Nevertheless, in view 
of the nonlinear nature of the defining
differential equations, they admit transseries
solutions and asymptotic expansions which 
can be used for accurate numerical
calculations. 
These results are useful in expansions
of partition and correlation functions
about instanton configurations~\cite{GiEtAl2024ii}.

Let us conclude by mentioning open problems,
which could inspire future research.
The first of these concerns the possibility 
of analytic expressions for the 
78-figure results reported here for 
particular function values and integrals
of the two- and three-dimensional instantons, 
notably, those communicated in Eqs.~\eqref{xiD20}---\eqref{integralsI}.
As already mentioned, we have performed 
a limited search based on the PSLQ 
algorithm~\cite{FeBa1992,BaPl1997,FeBaAr1999,BaBr2001}
using various Riemann zeta functions, 
logarithms, and polylogarithms, without finding 
suitable analytic formulas.
Our inability to find fully analytic representations
is mirrored in recent, somewhat related investigations~\cite{BoBr2022}.
The second open problem concerns the search 
for closed-form representations of the 
higher-order terms in the transseries 
solution~\eqref{uptoseven_3D} and~\eqref{uptoseven_2D},
generalizing the results
given in Eqs.~\eqref{compact3D} and~\eqref{compact2D}
to higher orders of the exponential factor $\exp(-r)$.

%
%
\section*{Acknowledgements}

The authors acknowledge insightful conversations with 
Professor Giorgio Parisi.
This work has been supported by the National Science Foundation (Grant
PHY--2110294) and by
the Swedish Research Council (Grant No.~638-2013-9243).  Support from the
Simons Foundation (Grant 454949) also is gratefully acknowledged. 
E.M.M.~acknowledges the MUR-Prin 2022 funding Prot.~20229T9EAT, financed by
the European Union (Next Generation EU).
 
\appendix

%
%
\section{Four--Dimensional Massless Quartic Theory}
\label{appa}

The existence of instantons in the renormalizable
(but not super-renormalizable) quartic 
theory in four dimensions has been anticipated
in Sec.~\ref{sec5A}. We consider the action
\begin{equation}
\calS[ \phi ] = \int \dd^4 x
\left[ \frac12 \left(\vec\nabla \phi \right)^2 +
\frac{1}{4} \, g \, \phi^4 \right] \,.
\end{equation}
The corresponding field equation is
$- \vec\nabla^2 \phi_\cl(\vec x) + g \, \phi_\cl^3(\vec x) = 0$.
We know that the solution of minimal action is spherically
symmetric, thus we  set
\begin{equation}
\phi_\cl(x) = \sqrt{-\frac{1}{g}} \, \xi_\cl(r) \,,
\end{equation}
where $r = | \vec x |$.
We then obtain a differential equation
$\left[ \left(\frac{ \dd }{ \dd r} \right)^2 +
\frac{3}{r} \frac{\dd}{\dd r} + \xi_\cl^2 (r) \right] \xi_\cl(r) = 0$.
The solution is 
\begin{equation}
\xi_\cl(r) = \frac{2 \sqrt{2}}{1 + r^2} 
\end{equation}
The instanton action is
\begin{equation}
S[\phi_\cl] = -\frac{A}{g} \,, 
\qquad A = \frac{8 \pi^2}{3} \,.
\end{equation}
The instanton integrals, $I_n = \int \dd^4 x \, \xi_\cl(r)^n$,
for $n = 3,4,6$, are
$I_3 = 8 \sqrt{2} \, \pi^2$,
$I_4 = \frac{32 \, \pi^2}{3}$,
$I_6 = \frac{128 \, \pi^2}{5}$.

%
%
\section{Six--Dimensional Massless Cubic Theory}
\label{appb}

Another example of the existence of 
analytically calculable instantons is the 
six-dimensional massless cubic theory~\cite{BoDuMe2021}.
We consider the action
\begin{equation}
\calS[ \phi ] = \int \dd^6 x
\left[ \frac12 \left(\vec\nabla \phi \right)^2 +
\frac{1}{3} \, g \, \phi^3 \right] \,.
\end{equation}
The corresponding field equation for the instanton
is $- \vec\nabla^2 \phi_\cl(\vec x) + g \, \phi_\cl^2(\vec x) = 0$.
We know that the solution of minimal action is spherically
symmetric, thus we  set
\begin{equation}
\phi_\cl(x) = \frac{1}{g} \, \xi_\cl(r) \,,
\end{equation}
where $r = | \vec x |$ and we observe that the 
instanton exists for positive $g$.
We then obtain a differential equation
$\left[ \left(\frac{ \dd }{ \dd r} \right)^2 +
\frac{5}{r} \frac{\dd}{\dd r} + \xi_\cl(r) \right] \xi_\cl(r) = 0$.
The solution is 
\begin{equation}
\xi_\cl(r) = -\frac{24}{(1 + r^2)^2} 
\end{equation}
The instanton action is
\begin{equation}
S[\phi_\cl] = \frac{A}{g^2} \,, 
\qquad A = \frac{192 \pi^3}{5} \,.
\end{equation}
The instanton integrals, $I_n = \int \dd^4 x \, \xi_\cl(r)^n$,
for $n = 3,4,6$, are $I_3 = -\frac{1152 \, \pi^3}{5}$,
$I_4 = \frac{55\,296 \, \pi^3}{35}$,
$I_6 = \frac{10\,616\,832\, \pi^3}{55}$.

\end{document}